\documentclass[aps,showpacs,pre,twocolumn,reprint]{revtex4-1}

\usepackage{graphicx,bm,amsmath,amssymb,natbib,url,epsfig}
\usepackage{dcolumn}
\usepackage{hyperref}

\usepackage{color}

\allowdisplaybreaks

\begin{document}

\title{Network resilience in the presence of non-equilibrium dynamics}


\author{Subhendu Bhandary${}^{1}$}
\author{Taranjot Kaur${}^{1}$}
\author{Tanmoy Banerjee${}^{2}$} \email{tbanerjee@phys.buruniv.ac.in}
\author{Partha Sharathi Dutta${}^{1}$}
\thanks{Corresponding author}
\email{parthasharathi@iitrpr.ac.in}

\affiliation{ ${}^{1}$Department of Mathematics, Indian Institute of
  Technology Ropar, Rupnagar 140 001, Punjab,
  India.\\ ${}^{2}$Chaos and Complex Systems Research
  Laboratory, Department of Physics, University of Burdwan, Burdwan
  713 104, West Bengal, India.}%

\received{:to be included by reviewer}
\date{\today}

\begin{abstract}
Many complex networks are known to exhibit sudden transitions between alternative steady states with contrasting properties. Such a sudden transition demonstrates a network's resilience, which is the ability of a system to persist in the face of perturbations. Most of the research on network resilience has focused on the transition from one equilibrium state to an alternative equilibrium state. Although the presence of non-equilibrium dynamics in some nodes may advance or delay sudden transitions in networks and give early warning signals of an impending collapse, it has not been studied much in the context of network resilience. Here we bridge this gap by studying a neuronal network model with diverse topologies, in which non-equilibrium dynamics may appear in the network even before the transition to a resting state from an active state in response to environmental stress deteriorating their external conditions. We find that the percentage of uncoupled nodes exhibiting non-equilibrium dynamics plays a vital role in determining the network's transition type. We show that a higher proportion of nodes with non-equilibrium dynamics can delay the tipping and increase networks' resilience against environmental stress, irrespective of their topology. Further, predictability of an upcoming transition weakens, as the network topology moves from regular to disordered. 
\end{abstract}

\maketitle

\section{Introduction}

Biological systems are often composed of nonlinear interactions and may exhibit complex multistable dynamics in response to various stressors, such as biochemical changes, environmental fluctuations, {\em etc} \cite{FePiSh18}. Empirical and theoretical evidences suggest that many multistable systems; including lakes \cite{scheffer2001catastrophic}, tropical forests \cite{HiHo11}, climate \cite{LeHeKr08,lenton2019climate}, cellular systems \cite{sharma2017regime,sarkar2019anticipating}, global finance \cite{MaLeSu08} can undergo sudden transitions to alternative stable states \cite{scheffer2009critical}. These transitions occur when a system bypasses a critical value, known as tipping points, due to the loss of resilience to perturbations \citep{walker2004resilience, folke2010resilience, saavedra2013estimating, LeverJ2014, barabasi2016network}. Determining such sudden transitions is relatively easy when a species or an isolated system governs the state of a system. However, for complex networked systems where many nodes interact, e.g., genetic networks, neural networks, and ecological networks of interacting species -- sudden transitions are highly unpredictable \cite{gao2016universal,liu2020network}. This is because current analytical frameworks of measuring resilience are mostly limited to low-dimensional systems and are underdeveloped for high-dimensional systems, where multiple components simultaneously interact through a complex network. Understanding, either the loss or the maintenance of network resilience in the face of internal disturbances or external environmental changes, is an important problem with broad interest.

Given a complex network, variations in the network topology can act as internal stress, which may perturb the system's resilience when exposed to stress deteriorating their external conditions (i.e., environmental stress) \citep{eom2018resilience}.  Investigations on how system parameters, such as growth rate and interaction strength, affect network resilience under changing environmental conditions, are primarily limited to mutualistic interactions \citep{saavedra2011strong,lever2014sudden}. However, real-world interactions are not restricted to mutualism but reveal diverse network architecture. Various social interactions, biochemical pathways, and protein-protein interactions primarily exhibit non-homogeneous properties that often vary in degree distribution with the power-law formation or the scale-free interactions.  Apart from diversity in network structure, the stability of complex networks is also accounted for by damages in the network parameters such as changes in the number of links or nodes, coupling strength, average degree, {\em etc}.  For example, \cite{albert2000error} studied the robustness of communication networks to loss of nodes in the disintegration of the network from a well-functioning state to a disturbed state. \cite{vazquez2003resilience} investigated the resilience to damage the networks in graphs with degree correlations.  Further, \cite{van2005implications, martin2015eluding} suggested that strength up to which a system can adapt changes to external conditions can increase by introducing substantial heterogeneity, noise, or weak coupling amongst the connected components. However, \citep{weissmann2018simulation} reported consideration of perfectly regular interaction to investigate network resilience \cite{martin2015eluding} leads to bias in the outcomes. Nevertheless, the investigations mentioned above mostly neglect the role of complex local dynamics, while investigating the interplay between external stress and network topology in the occurrence of tipping points.

While investigating the resilience patterns of complex networks, for mathematical tractability, it is appealing to consider either one-dimensional or two-dimensional nonlinear systems in each node with equilibrium steady state dynamics. However, this may overlook the concurrent effects of non-equilibrium dynamics preceding a transition \cite{tylianakis2014tipping}. Such non-equilibrium/cyclic dynamics may generate trade-off amongst other nodes, which can further suppress or fuel the occurrence of tipping points in the interaction network. In dynamical systems, cyclic dynamics mostly occur due to negative feedback, and positive feedback generates hysteresis loops, which creates the possibility of sudden transition of an equilibrium steady state. It would be interesting to study the resilience of a heterogeneous network, where a fraction of its nodes can exhibit equilibrium steady state, and a fraction of nodes can show cyclic dynamics. This new direction bear relevance for the management of tipping points in many networks, like neuronal and ecological networks, where oscillations are ubiquitous. 

Here we consider a network of FitzHugh-Nagumo (FHN) neurons \cite{fitzhugh1961impulses,Nagumo-1962}. The FHN model represents a paradigmatic neuronal model and have widely been used to study the collective behaviors of an ensemble of excitable as well as self excitatory spiking neurons. 
Several concepts of synchronization \cite{piko}, coherence resonance \cite{acanna1}, and chimera states \cite{ZAK20} have been discovered using networks of FHN neurons. Some of the emergent dynamics have been believed to have connections with observable physiological and cognitive behaviors. For example, the so-called bump state was proposed as the mechanism behind the visual orientation tuning in the rat's brain \cite{bard-neuron}. The unihemispheric slow-wave sleep in aquatic mammals and some migratory birds is connected with the coherent-incoherent dynamics of coupled oscillating neurons \cite{fhn-uni}. Recently, it has been shown explicitly that FHN oscillators on complex networks mimic epileptic-seizure-related synchronization phenomena \cite{fhn-seizer}.

In this work, one of our main interests is to study the influence of diverse network topologies and environmental stress on the FHN network resilience in the presence of non-equilibrium dynamics \cite{tylianakis2014tipping}.
We find that in a regular FHN network with heterogeneity in nodes, sometimes transitions from one stable state to another is gradual (i.e., continuous/second-order phase transition). At the same time, it can be sudden (i.e., discontinuous/first-order phase transition) for disordered networks. The management of internal parameters, such as coupling strength or average degree, can significantly influence the type of transitions. When the network is dominated by cyclic dynamics in proportion of nodes, an equilibrium network state precedes non-equilibrium dynamics. We find that a precise classification of transition type is not always possible. We observe fluctuations before a transition in both regular and disordered networks, under external stress, which indicates that the presence of non-equilibrium dynamics in a proportion of nodes plays a significant role in determining network resilience. Further, we find that the efficacy of covariance-based early warning signals weakens, as the network topology moves from regular to disordered.

\section{Mathematical model \label{s:model}}

\subsection{A single FHN neuron}

We consider the following model of the FHN neuron:
\begin{subequations}\label{eq1}
\begin{align}
 \frac{dx}{dt} &= \dfrac{1}{\epsilon}(x - \dfrac{1}{3}x^{3} - y), \label{eq1a}\\
 \frac{dy}{dt} &= gx - y + b \; ,
\end{align}
\end{subequations}
where $x$ is the membrane potential, and $y$ is the recovery
variable \Citep{shepelev2017bifurcations}. $\epsilon$ controls the time scale separation between the membrane potential ($x$) and the recovery variables ($y$) $(0<\epsilon<1)$\citep{fitzhugh1961impulses}. The parameters $b$ and $g$ are the control parameter of the neuron: depending upon them
each isolated FHN neuron can undergo several transitions between equilibrium and cyclic (or oscillating) dynamics. This transitions are governed by saddle-node bifurcation or/and Andronov-Hopf bifurcation.  

\begin{figure}
\centering \includegraphics[width=0.35\textwidth,angle=0, bb= 200 24 906 710]{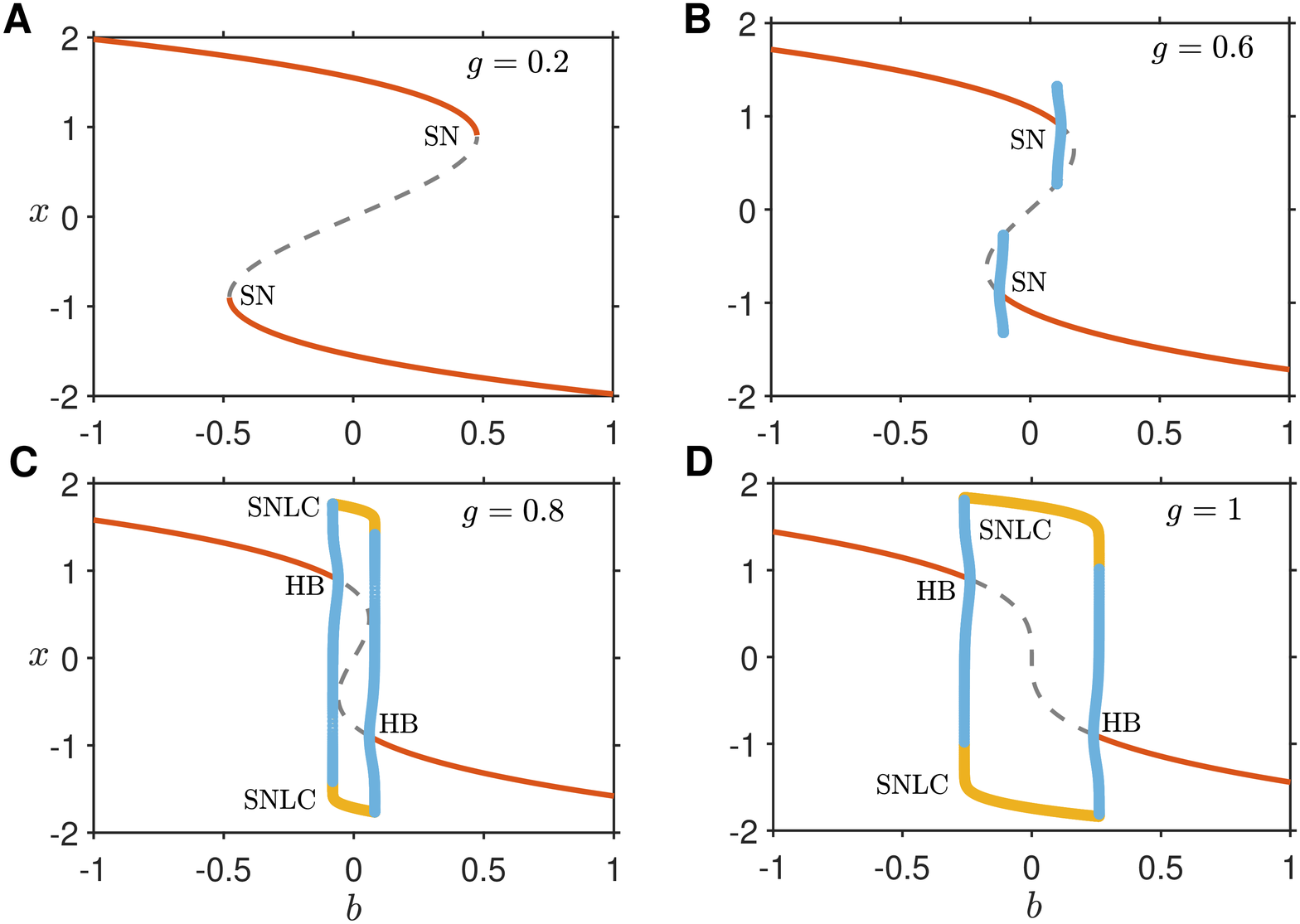}
\caption{One-parameter bifurcation diagram of the FHN equations~\eqref{eq1} along changing the parameter $b$: for (A) $g=0.2$, (B) $g=0.6$, (C) $g=0.8$, and (D) $g=1$. Solid lines and dashed lines represent stable and unstable steady states, respectively. Whereas yellow and blue filled circles represent stable and unstable limit cycles, respectively. Here, SN represents a saddle-node bifurcation point, HB represents a subcritical Andronov-Hopf bifurcation point, and SNLC represents a saddle-node bifurcation of the limit cycle. Other parameter value: $\epsilon=0.2$.}
\label{F1}
\end{figure}

For a clear understanding of a single neuron's underlying dynamics, bifurcation diagrams with $b$ for different values of $g$ are shown in Fig.~\ref{F1} (for $\epsilon =0.2$). When $g=0.2$, with variations in $b$ the model, shows a bistable region and forms a hysteresis loop governed by two saddle-node bifurcation points (Fig.~\ref{F1}A). For an increase in $g$, i.e.,~$g=0.6$, the stable branches lose their stability via a subcritical Andronov-Hopf bifurcation (Fig.~\ref{F1}B). With a further increase in $g$, there is a sudden transition from a steady state to a large amplitude limit cycle oscillations via a subcritical Andronov-Hopf bifurcation and saddle-node bifurcation of the limit cycle (Figs.~\ref{F1}C and~\ref{F1}D) with changes in $b$. The system experience sudden transitions from an equilibrium steady state to a non-equilibrium state (i.e., limit cycle) and vice versa. 

\subsection{Networks of FHN neuron}

Next, we consider a network of FHN neurons where dynamics of the $i$-th neuron in a network of $N$ neurons are governed as follows:
\begin{subequations}\label{eq2net}
\begin{align}
 \frac{dx_{i}}{dt} &= \dfrac{1}{\epsilon}(x_{i}-\dfrac{1}{3}x_{i}^{3}-y_{i})+\dfrac{d}{k_{i}} \sum_{j=1}^{N} A_{ij}(x_{j}-x_{i}),\\
 \frac{dy_{i}}{dt} &=g_{i} x_{i}-y_{i} + b, 
\end{align}
\end{subequations}
where $i=1, 2, \hdots, N$ denotes the node index and $d$ is the interaction strength between neurons. $A_{ij}$ is the element of the adjacency matrix ($A$) of the network, which is equal to $1$ if nodes $i$ and $j$ are connected and $0$ otherwise. $k_{i}$ denotes the number of connections corresponding to the $i$-th node $(k_i=\sum_{j=1}^{N}A_{ij})$. $g_{i}$ is the parameter through which heterogeneity among the nodes can be introduced in the network, and $b$ is considered as the external stress that drives the system to the transition point.

By simple algebraic manipulation we can rewrite \eqref{eq2net} in terms of the Laplacian matrix ($L$) as follows:
\begin{subequations}\label{eqNF}
\begin{align}
 \frac{dx_{i}}{dt} &= \dfrac{1}{\epsilon}(x_{i}-\dfrac{1}{3}x_{i}^{3}-y_{i})-\dfrac{d}{k_{i}} \sum_{j=1}^{N} L_{ij}
x_j, \\
 \frac{dy_{i}}{dt} &=g_{i} x_{i}-y_{i} + b,
\end{align}
\end{subequations}
where $L_{ij}$ is the element of the Laplacian matrix, such that $L_{ii}=\sum_{j=1}^{N}A_{ij}$ and $L_{ij}=-A_{ij}$ for $i \neq j$. Here we consider three different network architectures; the Watts-Strogatz (WS) model \citep{watts1998collective}, the Erd\H{o}s-R\'enyi (ER) model \citep{erdHos1960evolution}, and the Barab\'asi-Albert (BA) model \citep{barabasi1999emergence}. For the WS model, the interaction networks are generated using a rewiring probability ($p$). This probability determines the randomness of connectivity in a network. For example, the network is regular for $p = 0$ with each node corresponds to same degree distribution, completely random when $p = 1$, and is small-world for $0 < p < 1$. The ER and BA models also generate random networks, however one follows a Poisson distribution and another scale-free distribution, respectively. For all the aforementioned interaction networks, without any loss of generality, the average degree is kept at $k=\dfrac{1}{N} \sum_{i=1}^{N} k_{i} = 4$, with $k_i$ being degree of the $i$-th node; also, we consider $N=1000$ unless mentioned otherwise. 

\begin{figure*}
\centering \includegraphics[width=0.7\textwidth,angle=0, bb= 280 24 1250 700]{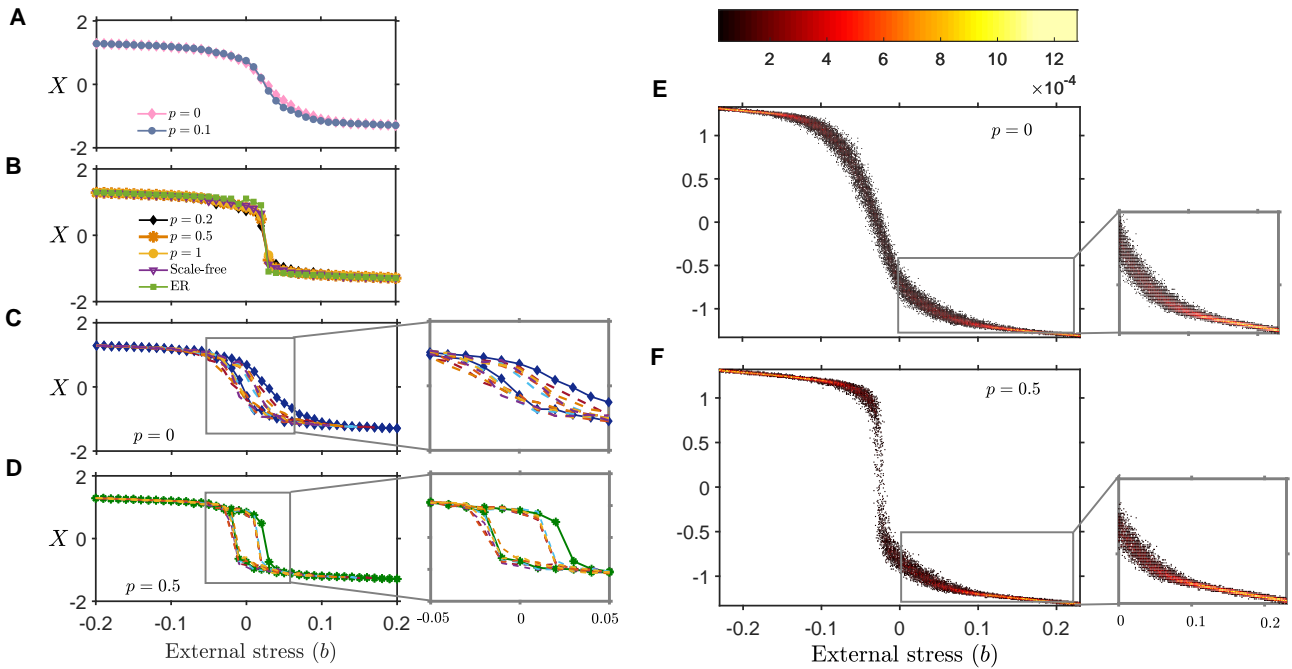}
\caption{Stationary state ($X$) of the FHN network with variations in $b$: for (A) regular ($p=0$) and small world ($p=0.1)$ networks, (B) small world topology with $p=0.2$ and p=$0.5$, WS, BA, and ER networks. (C-D) Each dashed line represents single realisation obtained by increasing the stress $b$ from $-0.3$ and then decreasing the stress $b$ once it reaches $b=0.07$, $b=0.09$, $b=0.11$, $b=0.13$, $b=0.15$, and $b=0.17$, for the WS network with $p=0$ and $p=0.5$. (E-F) Stationary state distribution of regular network ($p=0$), and small-world network ($p=0.5$), respectively. The blow-up diagrams represent  magnified dynamics observed for a range of $b$.  To estimate the stationary state distribution of a network we generate $5 \times 10^4$ stationary states, where $b$ is randomly selected from $[-0.2,0.2]$ following uniform distribution, similarly $g_{i}$ is randomly chosen from uniform distribution $[0.2,1]$, and initial conditions $(x_i,y_i)$ are randomly selected from $[-1.5,1.5] \times [0,1]$. The other parameters are $\epsilon =0.2$, and $d=1.5$.}
\label{F2}
\end{figure*}

\section{Results \label{s:res}}

For numerical integration, we use the 4th order Runge-Kutta method with adaptive step size and initial conditions $(x_i,y_i)$ are randomly selected from $[-1.5,1.5] \times [0,1]$. Throughout this paper we take $\epsilon =0.2$.

To analyze the resilience of the FHN networks, we calculate the stationary state of the network \eqref{eqNF}. We define the network state as $X=\frac{1}{N} \sum_{i=1}^{N}x_{i}$.  When $X >0$ (i.e., positive membrane potential), the network exhibits a potential difference between the cell membranes. Thus, signals transmitted between different parts of the cell and the neuronal interactions are active. When $X<0$ (i.e., negative membrane potential), the system is at the resting state.  Here, stationary states are calculated as an average of $100$ independent simulations.

\subsection{Role of network topology on determining the nature of transitions  \label{ss:1}}

We identify the role of network topology in response to the external stress ($b$) by obtaining the stationary network state $X$ for the WS, ER, and BA models. The stationary state $X$ is calculated by increasing the stress $b$ from $-0.3$ to $0.3$ with an increment of $ \delta b= 0.01$ and $g_{i}$ is randomly chosen from $[0.2,1]$ with a uniform distribution. We take the coupling strength $d=1.5$. Figure~\ref{F2} depicts the response of the network with different rewiring probability ($p$) of the WS model, varying from a perfectly regular network to a completely random network via the small-world networks. For low stresses, each network exhibits an active state while moving to a resting state as $b$ increases beyond a critical threshold. However, the nature of the transition from an active to a resting state depends upon the network topology. Regular networks reveal gradual response to the stress (Fig.~\ref{F2}A), which is a hallmark of second-order phase transition \Citep{van2005implications,martin2015eluding}. However, disordered networks (WS networks with $p\geq 0.2$, the ER and the BA networks) exhibit an abrupt transition from $X>0$ to $X<0$ (see Fig.~\ref{F2}B) a true discontinuous, first-order phase transition.

Further, we investigate the occurrence of hysteresis loops in the network when the external stress $b$ is decreased from $b=0.3$ to $b=-0.3$ (Figs.~\ref{F2}C--\ref{F2}D). We find that hysteresis loops occur for both the continuous and the discontinuous transitions. Each pathway of $X$ corresponds to increasing the stress $b$ from $-0.3$ and then decreasing it once the stress reaches to $b=0.07$, $b=0.09$, $b=0.11$, $b=0.13$, $b=0.15$, and $b=0.17$. Since all the pathways of transition from one state to another in both the scenarios are comparable, it suggests no strategy to mitigate the hysteresis in networks by controlling the external parameter $b$ in the WS networks (regular to disordered behavior). Thus, we find the transition threshold and changing environmental stress associated with tipping points in disordered networks, which can lead to a catastrophic transition. However, for regular networks, a transition from active to resting-state or vice-versa is not characterized by a tipping point.

To elucidate the nature of the transition thresholds, we generate stationary state distributions of \eqref{eqNF}, with different network topologies. A large number of stationary states are estimated to construct the stationary state distributions. We generate $5 \times 10^4$ stationary states ($X$'s), where values of $b$ are randomly selected from the interval $[-0.2,0.2]$ following the uniform distribution, initial conditions ($x_i,y_i$) are randomly selected from $[-1.5,1.5] \times [0,1]$, and $g_{i}$ is also randomly selected from $[0.2,1]$ following the uniform distribution. The stationary state distribution of the regular network differs qualitatively to the transition type viz a viz disordered networks (see Figs.~\ref{F2}E--\ref{F2}F). A higher probability of stationary states is distributed in the active ($X>0$) and resting ($X<0$) states for both the regular and the disordered networks. Within the transition regions between two states, stationary states are dense for regular networks (Fig.~\ref{F2}E), while it becomes scattered for the disordered networks (Fig.~\ref{F2}F). This implies that the transition in regular networks is continuous. Simultaneously, the stationary state distribution depicts a sudden jump from an upper to a lower state as randomness in network topology increases. Thus, regular networks have a higher degree of resilience to environmental perturbations than that of disordered networks. Overall, network topology significantly decides the nature of a system's transition in virtue of external stress. The state curve depicts gradual response for regular networks and abrupt change for disordered ($p \geq0.2$), scale-free, and ER networks.

\begin{figure}[!ht]
\centering \includegraphics[width=0.38\textwidth,angle=0,bb= 180 4 906 700]{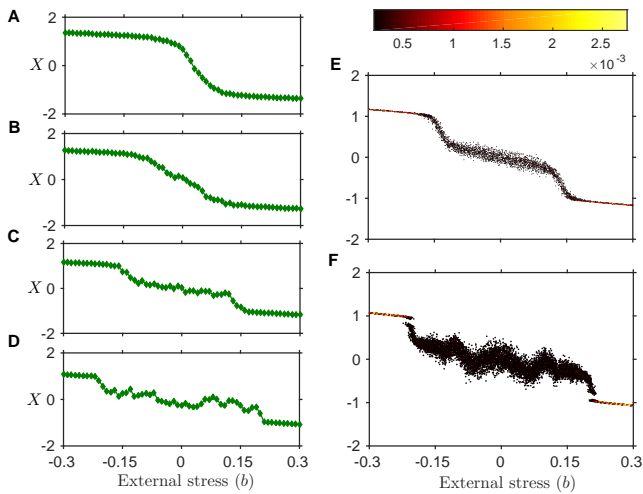}
\caption{Stationary states ($X$) of the regular network ($p=0$) with changes in $b$, where $g_i$'s are chosen from different intervals. $g_{i}$'s are randomly chosen, following the uniform distribution, from the intervals: (A) $g_{i} \in [0.2,1]$, (B) $g_{i} \in [0.4,1]$, (C) $g_{i} \in [0.6,1]$, and (D) $g_{i} \in [0.8,1]$. Stationary state distributions of the network when: (E) $g_{i}\in [0.6,1]$, and (F) $g_{i}\in [0.8,1]$. The initial conditions and other parameters are same as in Fig.~\ref{F2}.}
\label{F3}
\end{figure}

\begin{figure*}[!ht]
\centering 
\includegraphics[width=0.46\textwidth, angle=0,bb= 420 4 1000 700]{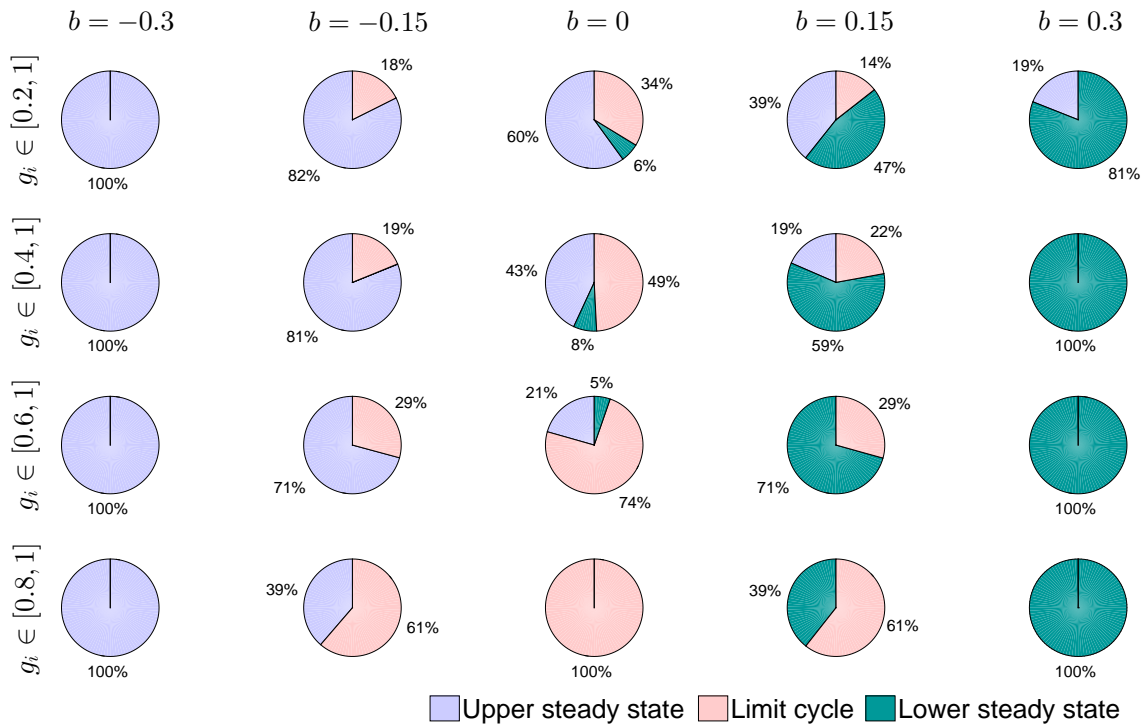}
\caption{Pie diagrams, when $d=0$, representing the percentage of nodes exhibiting steady states and cyclic dynamics, for different values of $b$ and range of heterogeneity $g_i$. We consider an ensemble average of $100$ realizations for each of the pie diagrams. }
\label{F3AddA}
\end{figure*}

\begin{figure}[h]
\centering \includegraphics[width=0.38\textwidth,angle=0,bb= 150 4 906 650]{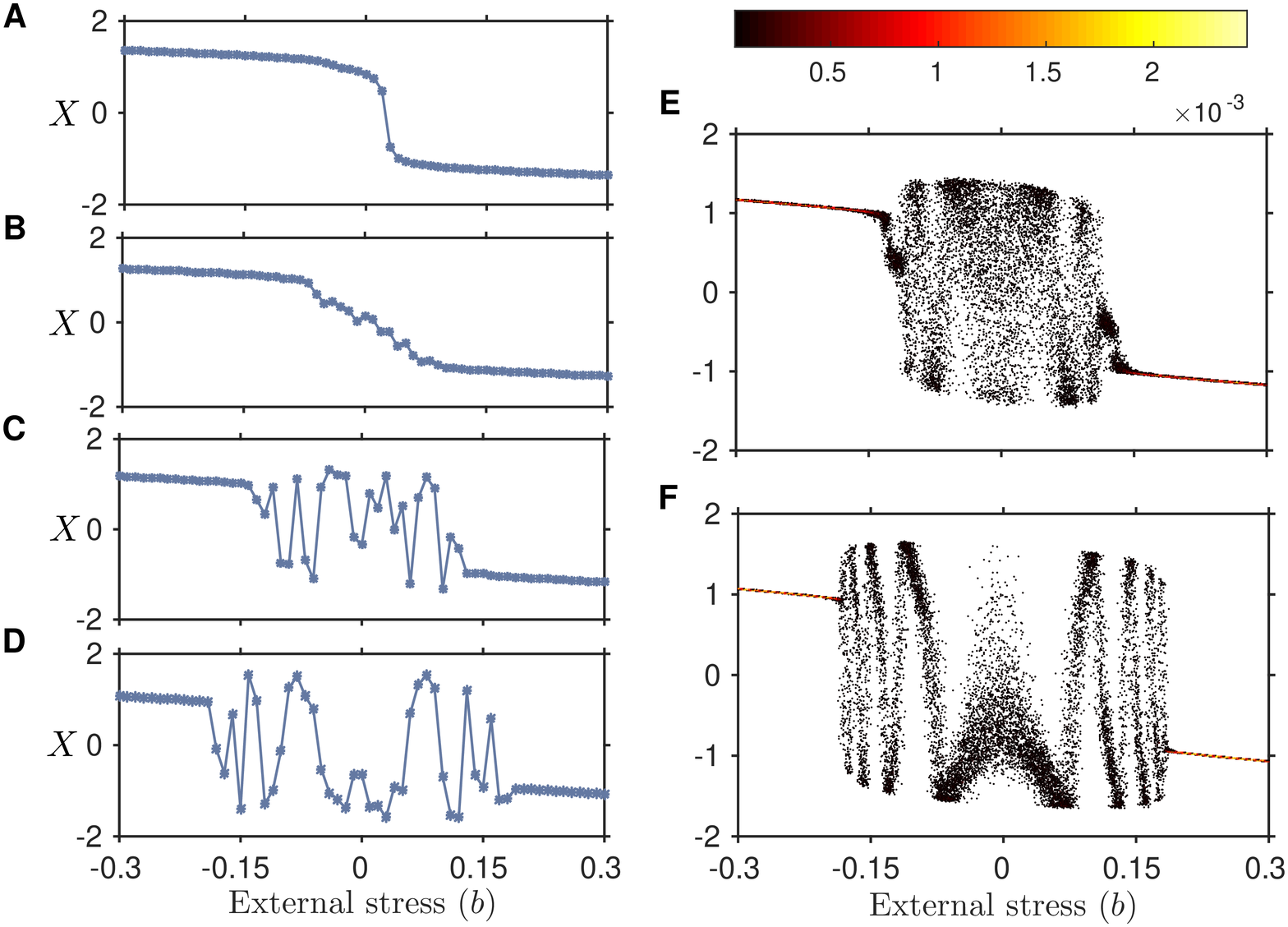}
\caption{Stationary states ($X$) of a disordered network ($p=0.5$) with changes in $b$, where $g_i$'s are randomly chosen, following the uniform distribution, from the intervals: (A) $g_{i} \in [0.2,1]$, (B) $g_{i} \in [0.4,1]$, (C) $g_{i} \in [0.6,1]$, and (D) $g_{i} \in [0.8,1]$. Stationary state distributions of the network when: (E) $g_{i}\in [0.6,1]$, and (F) $g_{i}\in [0.8,1]$. The initial conditions and other parameters are same as in Fig.~\ref{F2}.}
\label{F4}
\end{figure}

\subsection{Effects of node heterogeneity \label{ss:2}}

In this section, we study effects of different ranges of heterogeneity $g_i$ on the network resilience for diverse network topologies, when exposed to external stress $b$.  Figure \ref{F3} depicts the state curves for regular networks ($p=0$) considering different ranges of $g_{i}$. As the range of $g_i$ gets narrow (i.e., weak heterogeneity), we observe the significant influence of the limit cycle oscillations present in an isolated FHN neuron (see Figs.~\ref{F1}C--\ref{F1}D), during the transition in the network. When $g_i \in [0.2,1]$ (Fig.~\ref{F3}A) and $g_i \in [0.4,1]$ (Fig.~\ref{F3}B), there exists continuous transition like a transition from $X>0$ to $X<0$; however, the transition looks relatively flattered for the latter case. Further decrease in the heterogeneity range ($g_i \in [0.6,1]$ and $g_i \in [0.8,1]$) flattens the state curve near the transition threshold (Figs.~\ref{F3}C--\ref{F3}D) and it becomes difficult to decide the nature of the transition. Since the dynamics of a few uncoupled neurons have limit cycle oscillations, to identify the significance of the cyclic dynamics on the network, we calculate the distribution of the stationary states when $g_i \in [0.6,1]$ (Fig.~\ref{F3}E) and $g_i \in [0.8,1]$ (Fig.~\ref{F3}F). The stationary state distributions reveal a cluster of states in the vicinity of the transition thresholds. However, in the intermediate transition region, the stationary states get more scattered as the heterogeneity becomes weak (i.e., when we choose $g_i$ from a narrow interval). This suggests that the networks with weak heterogeneity in $g_i$ are more resilient as sudden transitions seem impossible. Therefore, the presence of limit cycle oscillations influences the network's resilience by inducing intermediate fluctuating stationary states as the network transits from an active to a resting state. Figure~\ref{F3AddA} depicts the proportion of nodes which are initially ($d=0$) either in upper steady state, or with cyclic dynamics, or in lower steady state with variations in the stress $b$ and the interval in which $g_i$ lies. The intermediate ranges of $b$, the $1000$ nodes are in different states due to the variations in $g_i$. It is evident that when the network reaches its stationary state, the proportion of node dynamics alters from its initial configuration.

Next, we analyze the impact of cyclic dynamics for the disordered small-world network ($p=0.5$). The stationary state curves depict that weakening of the heterogeneity parameter from $g_i \in [0.2,1]$ to $g_i \in [0.4,1]$ can suppress the sudden transition in the network and increase the resilience to the stress (see Figs.~\ref{F4}A--\ref{F4}B). Moreover, the cyclic dynamics' influence becomes more evident in the small-world network than the regular network. As $b$ moves from a negative to a positive value, we initially see that the state $X (>0)$ slowly approaching towards the transition point; however, the states are largely scattered before the transition to $X<0$ in the intermediate region. Here, large fluctuations between the two states characterize transitions (see Figs.~\ref{F4}C--\ref{F4}D). Thus, small-world network topology switches the network state between alternative states before it reaches a resting state. The stationary state distributions of the network also reveal a strong influence of the cyclic dynamics in inhibiting large fluctuations between the active and the resting state (Figs.~\ref{F4}E--\ref{F4}F). Hence, heterogeneity in $g_i$ can serve as a significant factor in determining the type of transition in the network.

\subsection{The role of network parameters \label{ss:3}}

\begin{figure*}[ht]
\centering \includegraphics[width=0.98\textwidth, angle=0]{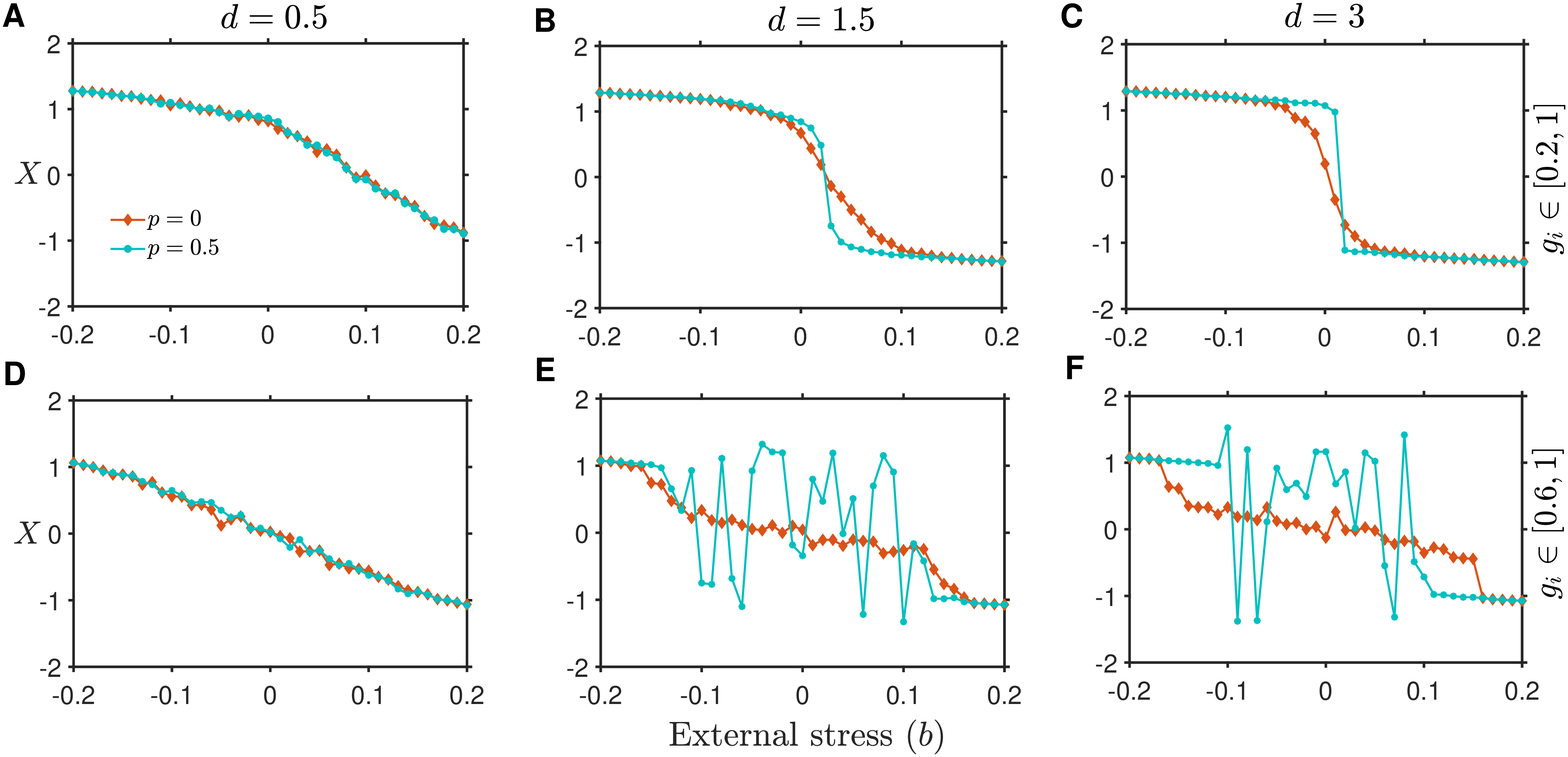}
\caption{The effect of different interaction strengths $d$ on network resilience for both the regular and disordered networks with changes in $b$: for (A-C) $g_i\in[0.2,1]$, and (D-F) $g_i\in[0.6,1]$. The influence of cyclic dynamics becomes evident when heterogeneity in $g_i$ becomes weak. The initial conditions and other parameters are same as in Fig.~\ref{F2}.}
\label{F5}
\end{figure*}

This section focuses on investigating the impact of network parameters, such as interaction strength ($d$) and average degree ($k$) on the resilience of the network \eqref{eqNF} to the stress $b$. We observe the influence of the network topology and the cyclic dynamics on the resilience patterns, in Subsections~\ref{ss:1}and \ref{ss:2}. Here, we calculate stationary state curves for different interaction strength ($d$), each for the regular and the disordered (small-world) networks as well as for different choice of the heterogeneity parameter $g_i$ (Fig.~\ref{F5}). In particular, to isolate the impact of cyclic dynamics and $d$ on the resilience of the network, we analyse the stationary state curves when $g_i \in [0.2,1]$ (Figs.~\ref{F5}A-\ref{F5}C), and $g_i \in [0.6,1]$ (Figs.~\ref{F5}D-\ref{F5}F). 

We find that the resilience of regular and small-world networks with substantial heterogeneity $(g_i\in[0.2,1]$) decreases as $d$ increases. For the weak interaction strength $d=0.5$, the type of transition is robust to network topology choice. In particular, we see continuous or second-order phase transitions (Fig.~\ref{F5}A). However, the network tends toward an abrupt transition as $d$ increases (Figs.~\ref{F5}A--\ref{F5}C). Therefore, in case of substantial heterogeneity in $g_i$, an increase in $d$ can reduce adaptive responses to stress and trigger sudden collapse to the resting state, irrespective of the network topology. Further, regular networks experience a relatively smooth transition to external stress for $g_i \in [0.6,1]$, while disordered networks show an increase in fluctuating states around transition thresholds with an increase in $d$. Therefore, when $g_i \in [0.6,1]$ for large enough $d$, it is not possible to determine the transition type of the stationary state. They are neither first-order nor second-order transition.

\begin{figure*}[ht]
\centering \includegraphics[width=0.98\textwidth, angle=0]{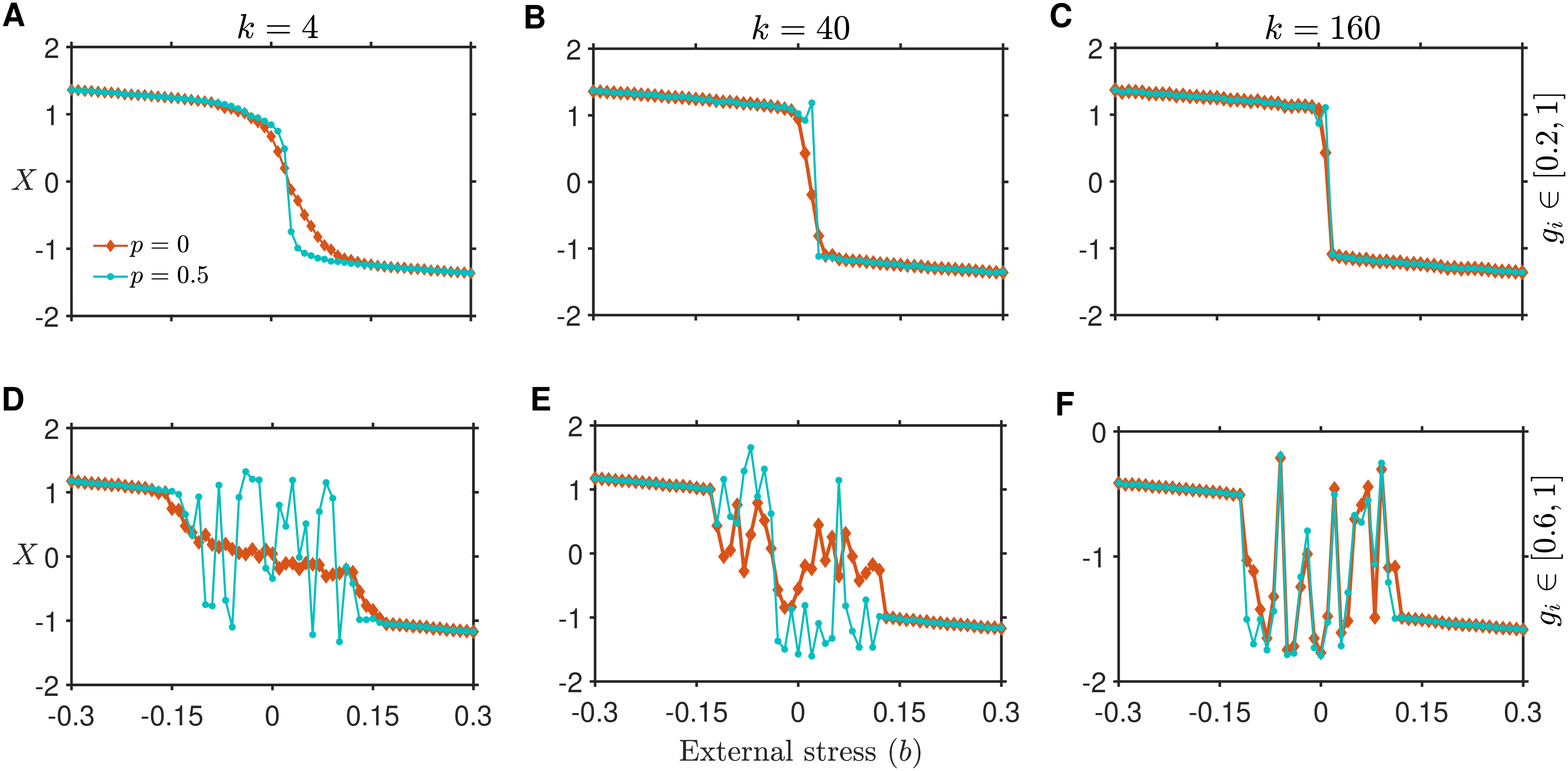}
\caption{The effect of different average degree $k$ on network resilience for both the regular and disordered networks with changes in $b$: for (A-C) $g_i\in[0.2,1]$, and (D-F) $g_i\in[0.6,1]$. The influence of cyclic dynamics becomes evident when heterogeneity in $g_i$ becomes weak. The initial conditions and other parameters are same as in Fig.~\ref{F2}.} \label{F6}
\end{figure*}

Similarly, an increase in the network's average degree $k$ leads to more stronger abrupt transition from one state to another for the system with strong heterogeneity ($g_i \in [0.2,1]$), in case of both $p=0$ and $p=0.5$ (Figs.~\ref{F6}A--\ref{F6}C). Note that changes in the network topology do not alter the system's response to $b$ for high average degrees and strong heterogeneity in the system. For $g_i \in [0.6,1]$, increase in $k$ for both the network topologies leads to strong fluctuations prior to a transition to a resting state (Figs.~\ref{F6}D--\ref{F6}F). Moreover, the fluctuations observed here for $p=0.5$ are relatively stronger than those observed for $p=0$ ( Figs.~\ref{F5}D--\ref{F5}F). Therefore, an increase in $k$ and incorporating weak heterogeneity makes it difficult to determine the transition type. Overall, strong heterogeneity in $g$ can improve networks' resilience with a low average degree and low coupling strength, while weak heterogeneity increases networks' resilience with strong $d$ and $k$.

\begin{figure}[ht]
\centering \includegraphics[width=0.4\textwidth, angle=0,bb= 210 0 1200 700]{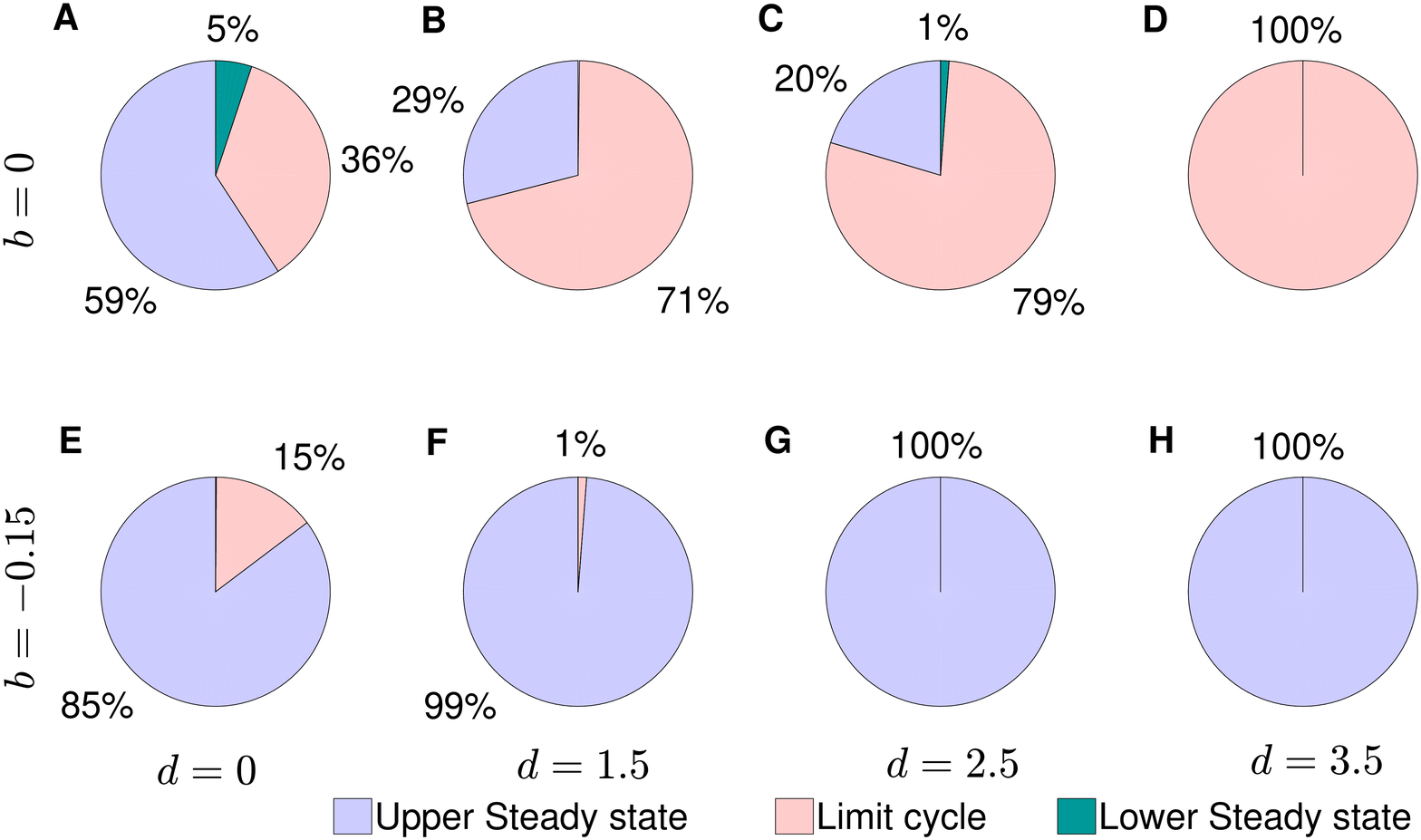}
\caption{Pie diagrams representing the percentage of nodes exhibiting steady state and cyclic dynamics, for different values of $d$ in a regular network: for (A-D) $b=0$, and (E-H) $b=-0.15$. The initial conditions ($x_i,y_i$) are randomly chosen following uniform distribution from $[-1.5,1.5]\times[0,1]$ with $d=0$, and then we calculate the proportion of nodes exhibiting steady state and oscillatory dynamics after removing the transients. For increasing values of $d$, the initial conditions used are same as for the system with no coupling ($d=0$). We consider ensemble average of $100$ realisations for each of the pie diagram.}
\label{F8}
\end{figure}

To check the stationary state of each node $x_i$, $i=1,2, \hdots, 1000$ for different interaction strengths $d$ and external stress $b$, we calculate the percentage of nodes those exhibit either steady state or cyclic dynamics (see Fig.~\ref{F8}). When the stress is close to the transition threshold ($b=0$), we observe that for each isolated node ($d=0$), the system is dominated by steady state dynamics (Fig.~\ref{F8}A). An increase in $d$ leads to an oscillatory state in a large number of nodes (Figs.~\ref{F8}B--\ref{F8}C), which further shows only oscillatory dynamics in all the nodes as $d$ becomes very high (Fig.~\ref{F8}D). Hence, when $d$ is low, some nodes exhibit steady state behavior while others fluctuate between active and resting states. The above results can lead to two exciting outcomes: First, if the steady state of the nodes represents their presence at an upper attractor, a large number of nodes exhibit functional stability. An increase in $d$ decreases the network's functional stability as all the nodes have changed their state from an upper attractor to an oscillatory state between the two attractors. Similarly, suppose the steady state for $d=0$ represents the nodes are at resting state. In that case, an increase in $d$ will enhance the functional stability as many nodes have shifted from a completely resting state to fluctuate between active and resting states.

The outcome is different for a negative value of the stress ($b=-0.15$), which is away from the transition threshold (Fig.~\ref{F8}). Initially ($d=0$), steady states are dominating a majority of the nodes. As $d$ increases, the percentage of cyclic states reduce, and finally, for large enough $d$ dynamics in all the nodes become a steady state (Figs.~\ref{F8}E--\ref{F8}G).

\subsection{Role of network topology on predictability of state transition}

Analyses of diverse network topologies have shown different system responses while exhibiting transition from an active to a resting state. The regular network reveals a gradual transition of the FHN neurons network from an active state, transmitting signals to different parts of cells, to a resting state. However, this transition is sometimes discontinuous, giving rise to the tipping point in small-world interactions. Also, we observe fluctuating dynamics in the system with weak heterogeneity before an actual transition takes place.

In many systems, a transition from a stable state to an alternative state preceded by a decrease in the equilibrium state's resilience. Studies suggest various statistical methods due to critical slowing down to anticipate transitions that occur in natural systems. In spatial models, critical slowing down is preceded by an increase in the state variable's spatial correlation and can serve as an early warning signal. Hence, it leads to an increase in fluctuations in the state variable, influencing fluctuations in the nearby state variables, leading to a more significant correlation in neighbouring components. Here, we investigate the impact of network architecture; in other words, different transition types on the early warning indicator of spatial systems.

Consider a general form of a deterministic framework followed by stochastic perturbations \citep{wilkinson2009stochastic,gammaitoni1995stochastic}: 
\begin{equation}\label{eq3}
d\textbf{z}(t)=f(\textbf{z}(t),b)dt+DdW(t),
 \end{equation}
where deterministic part of the system is expressed as a vector function $\textbf{f}=\{f_{1},f_{2},f_{3},\dots,f_{N}\}$ and state of the system $\textbf{z}(t) = \{z_{1}(t), z_{2}(t), \dots, z_{N}(t)\} $ is controlled by the dynamics of \eqref{eq3}. $W(t)$ is a Wiener process with $D$ as the noise intensity. Consider a vector of small deviation in the state $\textbf{u}=\textbf{z}-\textbf{z}^{*}$, where $\textbf{z}^{*}$ is the equilibrium point in the absence of any perturbation. Local stability of the fixed point $\textbf{z}^{*}$ is calculated from the linearisation of the perturbed differential equation:
\begin{equation}
 \frac{d\textbf{z}}{dt}=\frac{d\textbf{u}}{dt}=\textbf{f}(\textbf{z})=\textbf{f}(\textbf{z}
 ^{*}+\textbf{u})\approx \textbf{f}(\textbf{z}^{*})+J\textbf{u} = J\textbf{u},
 \end{equation}
where $J$ is the Jacobian matrix that determines the expected trajectory of the perturbed state as it returns to its equilibrium state $z^*$. Consider a probability distribution function $\rho(\textbf{u},t)$ to be a solution of \eqref{eq3} which can be approximated as a solution of the below linear
 Fokker Planck equation \Citep{risken1996fokker}:
 \begin{equation}
 \frac{\partial \rho(\textbf{u},t)}{\partial t}= -\sum_{i,j=1}^{N} J_{ij} \frac{\partial (\rho u_{i})}{\partial u_{i}}+\frac{1}{2}\sum_{i,j=1}^{N} D_{ij} \frac{\partial^2 \rho}{\partial^2 u_{i}}.
 \end{equation} 
The correlation between two state variables can be determined by a covariance matrix, where the diagonal elements represent fluctuations at each node and off-diagonal entries represent fluctuations between pair of nodes. The covariance matrix $C$ can be written as \Citep{kwon2005structure}: 
 \begin{equation}\label{eq6}
 C=-J^{-1}(D+Q)/2,
 \end{equation}
 where $Q$ is an antisymmetric matrix that can be estimated as:
 \begin{equation}\label{eq7}
 JQ+QJ^{T}=JD-DJ^{T},
 \end{equation}
where the superscript $T$ denotes the transpose of a matrix.

Near the bifurcation/tipping threshold, the dominant eigenvalue of the Jacobian matrix $J$ tends to zero, and the eigenvector corresponding to the dominant eigenvalue will become slower. The rate at which the covariance matrix's dominant eigenvalue converges to zero is relatively faster than those of the other eigenvalues. Thus, the dominant eigenvalue and its ratio over the Euclidean norm of a vector consisting of all eigenvalues of the covariance matrix serve as an early warning indicator in spatial interactions \Citep{chen2019eigenvalues}.

\begin{figure*}[!ht]
\centering \includegraphics[width=0.85\textwidth]{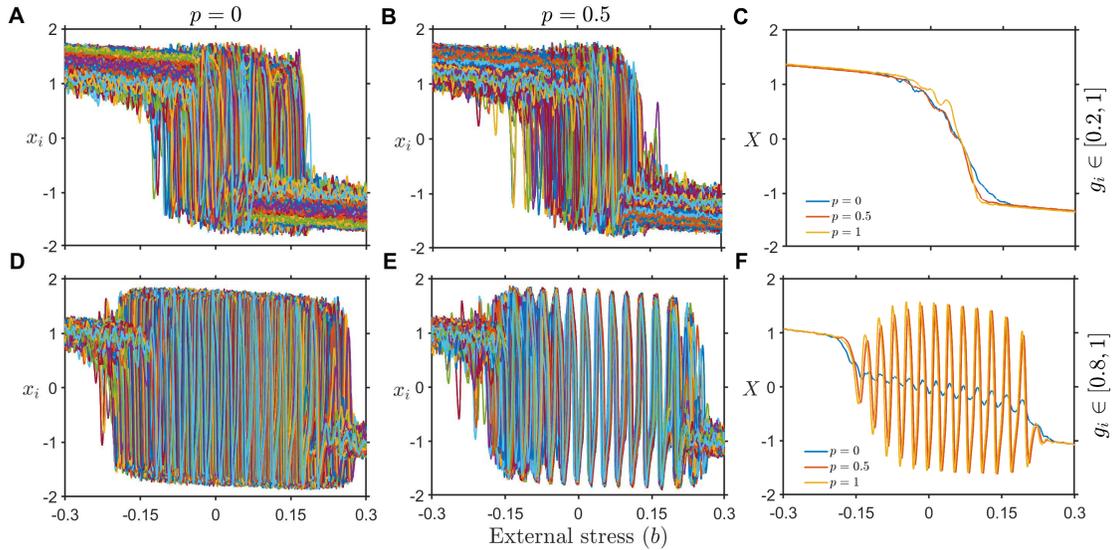}
\caption{Stochastic time series of the FHN neuron model (Eq.~\ref{eq8}) for different network topology and heterogeneity in $g$. (A-B) Each line represents the time series of the potential variable at each node, when the network follows strong heterogeneity in $g$ and has perfectly regular and small-world topology $(p=0.5$), respectively. (C) Stationary state time series for the WS network with $p=0$, $p=0.5$ and $p=1$. (D)-(E) stochastic time series of each node when the heterogeneity become weak, that is, $g_i$ is chosen randomly from uniform distribution $[0.8,1]$. (F) stochastic time series of the stationary state of the interaction network with weak heterogeneity.}
\label{F9}
\end{figure*}

\begin{figure}[!ht]
\centering \includegraphics[width=0.48\textwidth,angle=0]{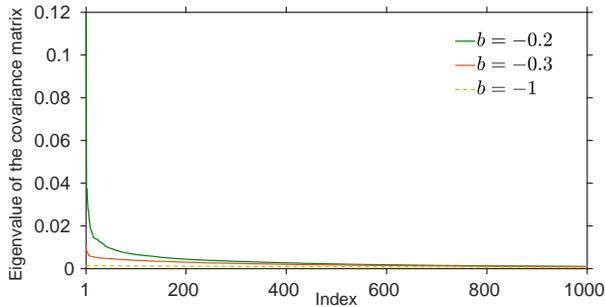}
\caption{ Spectrum of the covariance matrix as the system approaches close to the transition threshold. Each line represents the eigenvalues of the covariance matrix changes for different parametric values of the stress $b$. The Index ${i}$ represents the index of $i^{th}$ eigenvalue corresponding to the potential variable of the network, when each eigenvalue is arranged in descending order.
} \label{F11}
\end{figure}

\begin{figure}[!ht]
\centering \includegraphics[width=0.56\textwidth,angle=0,bb= 30 24 1106 750]{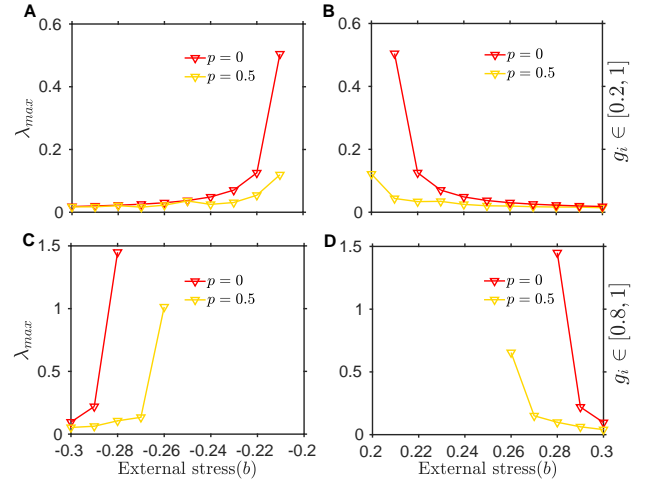}
\caption{ Largest eigenvalue $\lambda_{max}$ of the covariance matrix along with changes in the parameter $b$ for two different network structure depending on rewiring probability. (A-B) $g_{i}$ is randomly selected from uniform distribution $[0.2,1]$. (C-D) $g_{i}$ is randomly selected from uniform distribution $[0.8,1]$. }
\label{F10}
\end{figure}


We begin our investigation by considering \textbf{f} as the deterministic FHN network~\eqref{eqNF} with $N$ nodes. Each node has $\textbf{z}=\{x_{1}, y_{1}, x_{2}, y_{2}, \dots x_{N}, y_{N}\}$ as the pair-wise potential and recovery variables, respectively. Therefore, \eqref{eq3} can be expressed as:
\begin{subequations}\label{eq8}
\begin{align}
 {dz_{2i-1}}={dx_{i}} &= \left(\dfrac{1}{\epsilon}(x_{i}-\dfrac{x_{i}^{3}}{3}-y_{i})-\dfrac{d}{k_{i}} \sum_{j=1}^{N} L_{ij}
x_j\right)dt \nonumber\\ & +\sigma dW_{i}, \label{eq8a}\\
 {dz_{2i}}={dy_{i}} &=(g_{i} x_{i}-y_{i} + b)dt+\sigma dW_{i},  
\end{align}
\end{subequations}
where, $\sigma I = D$, and $\sigma=0.01$ represents the noise intensity of each node. We generate the stochastic time series of the FHN network along changing external stress $b$, having both strong (Figs.~\ref{F9}A--\ref{F9}C) as well as weak heterogeneity (Figs.~\ref{F9}D--\ref{F9}F) in $g_i$, using \eqref{eq8}. The stochastic time series along changing external stress, for each neuron ($x_i$) (Figs.~\ref{F9}A--\ref{F9}B) reveals transition from active to resting state. Furthermore, the stationary state ($X$) stochastic time series reveals that this transition is relatively gradual for low rewiring probability of the network (Fig.~\ref{F9}C). Investigating the stochastic time series for network with weak heterogeneity ($g_i \in [0.8,1]$) depicts significant role of oscillatory dynamics in the state of each node (Figs.~\ref{F9}D--\ref{F9}E). We observe that the oscillation exhibit high amplitudes and demonstrate large fluctuations between alternative stable states for the small-world to completely random network. However, smaller fluctuations are identified when the network is perfectly regular (Fig.~\ref{F9}F). Further, we calculate eigenvalues of the covariance matrix (using Eq.~\ref{eq6} and Eq.~\ref{eq7}) of the stochastic system, where the Jacobian matrix $J$ is given by: 
\begin{subequations}
\begin{align}
 J_{2i-1,2j-1}=
 \begin{cases}
 \frac{1}{\epsilon}(1-x_{i}^2)-\frac{d}{k_{i}}L_{ij}|_{z=z^{*}},~~ \text{if} ~~ i=j, \label{eq10}\\
 \frac{d}{k_{i}},~~ \text{if} \ i \neq j,~\text{and}~i,j~\text{are connected,}\\
0, ~ \text{ if } i \neq j,~\text{and}~i,j~\text{are not connected,}\\
 \end{cases} \nonumber
 \end{align}
\end{subequations}
\begin{subequations}\label{eq10a}
\begin{align}
J_{2i-1,2j}=
\begin{cases}
-1, & \text{if}\ i = j,\\
0, & \text{if}\ i \neq j,\\ 
\end{cases}\nonumber
\end{align}
\end{subequations}
\begin{subequations}\label{eq11}
\begin{align}
J_{2i,2j-1}=
\begin{cases}
g_{i}, & \text{if}\ i = j,\\
0, & \text{if}\ i \neq j,\\ 
\end{cases}\nonumber
\end{align}
\end{subequations}
\begin{subequations}\label{eq12}
and
\begin{align}
J_{2i,2j}=
\begin{cases}
-1, & \text{if}\ i = j,\\
0, & \text{if}\ i \neq j\;.\\ 
\end{cases}\nonumber
\end{align}
\end{subequations}
Let $\lambda_{max}=\lambda_1 \leq \lambda_2 \leq  \hdots \leq \lambda_{1000}=\lambda_{min}$ be the eigenvalues corresponding to the potential variable of the covariance matrix $C$. Index is the integer varying from $1$ to $1000$ such that Index-$1$ denotes the index of largest eigenvalue, Index-2 corresponds to the index of second largest eigenvalue, and so on. We observe that the dominant eigenvalue (Index-$1$) of $C$ increases at a larger rate than the other eigenvalues. Also, the difference between the dominant eigenvalue and other eigenvalues increases as the system approaches close to the transition threshold (Fig.~\ref{F11}). Thus, the dominant eigenvalue serves as an early warning indicator for the network. To investigate the role of different network topologies in predictability of the transition, we analyse the trend in the dominant eigenvalue of the covariance matrix along with the increasing/decreasing the stress $b$ (Fig.~\ref{F10}). We observe that as the network approaches in the vicinity of transition point, the dominant eigenvalue shows an increasing trend (Figs.~\ref{F10}A--\ref{F10}B). In case of weak heterogeneity, an increase in the largest eigenvalue is observed at relatively low value of $b$ (Figs.~\ref{F10}C--\ref{F10}D). This is due to fluctuations preceded by the state of the system before the transition happens. Importantly, we notice that the increase is relatively higher for the regular network than that for the small world interaction. This suggests that the strength of predictability of an upcoming transition depends upon the network topology.

\section{Conclusions and Discussion \label{s:dis}}

Dynamical transitions from one stable state to another alternative stable state are hallmarks of loss of resilience in many complex systems \citep{perrings1998resilience, walker2004resilience,folke2010resilience}. Studies provide a myriad of mechanisms that lead to the occurrence of such transitions. These mechanisms propel the transition to an alternative stable state when an input condition by-pass a critical point \citep{may1977thresholds, rietkerk2004self,HiHo11}. For instance, change in population growth due to reduced population size, changes in interaction strength between spatially connected components, {\em etc}. However, it has been challenging to understand how mechanisms altering the static dynamics of complex networks affect the occurrence of catastrophic and non-catastrophic transitions. Here, we aim to uncover, together with the presence of equilibrium dynamics, how non-equilibrium/cyclic dynamics in a few nodes of a network influence its resilience. We find that the proportion of uncoupled nodes exhibiting non-equilibrium dynamics plays a crucial role in determining the transition type of the network. A higher proportion of nodes with non-equilibrium dynamics can delay the tipping by inducing flickering between alternative network states against environmental stress, irrespective of their topology. Moreover, the predictability of a forthcoming transition weakens, as the network topology moves from regular to disordered.

Real-world disordered networks undergo deteriorated functional stability, often exhibiting discontinuous or first-order phase transition to an alternative state, as the input conditions change. Such sudden transitions in a network occur due to the inability of many nodes to cope with internal or external stresses. On the other hand, regular networks can buffer the collapse of a network state in response to increased environmental stress, overall supporting the network beyond a tipping point. Notably, the network's internal factors, such as average degree and coupling strength, can also regulate an abrupt transition. For weak interaction strengths, the change in network topology does not impact the system's resilience, while it can delay or remove the discontinuous transition in the system's state. The strong coupling can decrease the resilience of both regular as well as disordered networks. In contrast, disordered networks are robust to change in the network's average degree, as they reveal a sudden transition to an alternative state at low and high network degrees. We observe that increasing network degree decreases the resilience of the regular networks, and transitions become abrupt.

Furthermore, we find that each node's dynamics can significantly influence the response of a network's state with variations in external environmental stress. The dominance of limit cycle oscillations in the isolated components can shape the interaction network's response to changing stress. When cyclic dynamics dominate, increased coupling strength buffers the sudden transition by the emergence of fluctuations before transitioning to an alternate state. The propensity of fluctuations, however, is observed to be more for disordered networks. Thus, cyclic dynamics significantly delay the onset of a system's sudden transition to a collapsed state, and the resilience of disordered networks can be enhanced by controlling the coupling parameter and the network's average degree. We find that entirely random, ER, and BA networks share similar outcomes observed for disordered networks with small-world topology. Overall, the presence of non-equilibrium dynamics in nodes can improve network resilience by delaying sudden transitions and flickering between alternative states in the network preceding a transition can also work as an early warning signal \cite{wang2012flickering}.

While some real networks are highly resilient to environmental stress, others show vulnerable response towards it. The nature and scale of sudden transitions in a network due to changing environmental stress can significantly owe to  the diversity of network topology, internal perturbations, and external heterogeneity. Hence, we analyze early warning signals \citep{Dakos:2012pone,Scheffer:2012sc} to foresee such varying nature of transitions. We test the predictability of network-based early warning indicators for both the regular and the small-world topology \citep{chen2019eigenvalues} by investigating the covariance matrix of our system in the presence of additive Gaussian white noise. We observe that in the vicinity of the transition phase, the variance of the fluctuations increases. This increase is more substantial for regular than that for the small-world networks. While many real-world networks may undergo an abrupt transition from one stable state to an alternative stable state, such transitions' predictability depends upon the network topology. Finally, understanding network resilience and forecasting sudden transitions in the face of rapidly evolving stress can be considered a challenging future problem. 

\begin{acknowledgments}
P.S.D. acknowledges financial support from the Science \& Engineering Research Board (SERB), Govt.~of~India [Grant No.: CRG/2019/002402].
\end{acknowledgments}


\end{document}